\documentclass[12pt]{iopart}

\usepackage{amssymb}
\usepackage{graphicx}
\usepackage{color}

\begin{document}

\title[Neutrino cosmology and PLANCK]{Neutrino cosmology and PLANCK }

\author{Julien Lesgourgues$^1$ and Sergio Pastor$^2$}

\address{$^1$ Institut de Th{\'e}orie des Ph{\'e}nom{\`e}nes Physiques, EPFL, CH-1015 Lausanne, Switzerland\\ 
 	 CERN, Theory Division, CH-1211 Geneva 23, Switzerland\\
          LAPTh (CNRS-Universit{\'e} de Savoie), B.P.\ 110, F-74941 Annecy-le-Vieux Cedex, France}
                
\address{$^2$ Instituto de F\'{\i}sica Corpuscular  (CSIC-Universitat de Val\`{e}ncia),
c/ Catedr\'atico Jos\'e Beltr\'an, 2, 46980 Paterna (Valencia), Spain}
\eads{\mailto{Julien.Lesgourgues@cern.ch}, \mailto{Sergio.Pastor@ific.uv.es}}
%\ead{custserv@iop.org}
\begin{abstract}
Relic neutrinos play an important role in the evolution of the Universe,
modifying some of the cosmological observables. We summarize the main 
aspects of cosmological neutrinos and 
describe how the precision of present cosmological data can be used to
learn about neutrino properties.
In particular, we discuss how cosmology provides information on the absolute scale
of neutrino masses, complementary to beta decay and neutrinoless double-beta
decay experiments. We explain why the combination of Planck temperature data with measurements of the baryon acoustic oscillation angular scale provides a strong bound on the sum of neutrino masses, 0.23~eV at the 95\% confidence level, while the lensing potential spectrum and the cluster mass function measured by Planck are compatible with larger values. 
We also review the constraints from current data on other neutrino properties.
Finally, we describe the very good perspectives
from future cosmological measurements, which are expected to be
sensitive to neutrino masses close the minimum values 
guaranteed by flavour oscillations.
\end{abstract}

\submitto{\NJP}
\maketitle

\section{Introduction}
\label{20-sec:intro}

The role of neutrinos in
cosmology is one of the best examples of the very close ties that have
developed between nuclear physics, particle physics, astrophysics and
cosmology. Here we focus on the most interesting aspects related to 
the case of massive (and light) relic neutrinos, but many
others that were left out can be found in specialised books \cite{20-NuCosmo} and 
reviews \cite{20-Dolgov:2002wy,20-Hannestad:2006zg,20-Lesgourgues:2006nd,20-Hannestad:2010kz,20-Wong:2011ip,Archidiacono:2013fha}.

We begin with a description of the properties and evolution of the
background of relic neutrinos that fills the Universe. 
The largest part of this paper is devoted to the impact of massive
neutrinos on cosmological observables, that can be used to extract
bounds on neutrino masses from present data, with emphasis on the
results of the Planck satellite. Next, we review the implication of current cosmological data on other neutrino properties (relic density, leptonic asymmetry, extra sterile neutrino species). Finally, we discuss the
sensitivities of future cosmological experiments to neutrino masses.
%
%A more general review on the connection between particle physics and cosmology can be found in
%\cite{20-Kamionkowski:1999qc}.

\section{The cosmic neutrino background}
\label{20-sec:theCNB}

The existence of a relic sea of neutrinos is a generic feature of the
standard hot big bang model, in number only slightly below that of
relic photons that constitute the cosmic microwave background
(CMB). This cosmic neutrino background (CNB) has not been detected
yet, but its presence is indirectly established by the accurate
agreement between the calculated and observed primordial abundances of
light elements, as well as from the analysis of the power spectrum of
CMB anisotropies and other cosmological observables.  
%In this section we will summarize the evolution and main properties of the CNB.

%\subsection{Relic neutrino production and decoupling}
%\label{20-subsec:nudec}

Produced at large temperatures by frequent weak interactions, flavour
neutrinos ($\nu_{e,\mu,\tau}$) were kept in
equilibrium until these processes became ineffective in the course of
the expansion of the early Universe. While coupled to the rest of the
primeval plasma (relativistic particles such as electrons, positrons
and photons), neutrinos had a momentum spectrum with an equilibrium
Fermi-Dirac form with temperature $T$,
\begin{equation}
f_{\rm eq}(p,T)=\left
[\exp\left(\frac{p-\mu_\nu}{T}\right)+1\right]^{-1}\,,
\label{20-FD}
\end{equation}
which is just one example of the general case of particles in
equilibrium (fermions or bosons, relativistic or non-relativistic), as
shown e.g.\ in \cite{20-kt}. In the previous equation we have included a
neutrino chemical potential $\mu_\nu$ that would exist in the presence
of a neutrino-antineutrino asymmetry, but it has been shown 
that even if it exists its contribution can not be very relevant
\cite{20-Mangano:2011ip}.

As the Universe cools, the weak interaction rate falls
below the expansion rate and neutrinos decouple from the
rest of the plasma. An estimate of the decoupling temperature $T_{\rm
dec}$ can be found by equating the thermally averaged value of the
weak interaction rate
$\Gamma_\nu=\langle\sigma_\nu\,n_\nu\rangle$,
where $\sigma_\nu \propto G_F^2$ is the cross section of the
electron-neutrino processes with $G_F$ the Fermi constant and
$n_\nu$ is the neutrino number density, with the expansion rate
given by the Hubble parameter $H 
=(8\pi\rho/3M_P^2)^{1/2}$.
Here $\rho\propto T^4$ is the total energy density, dominated by radiation, and $M_P=1/G^{1/2}$ is the Planck mass.  If we
approximate the numerical factors to unity, with $\Gamma_\nu \approx
G_F^2T^5$ and $H \approx T^2/M_P$, we obtain the rough estimate
$T_{\rm dec}\approx 1$ MeV.  More accurate calculations give slightly
higher values of $T_{\rm dec}$ which are flavour dependent because
electron neutrinos and antineutrinos are in closer contact with $e^\pm$, as shown e.g.\ in \cite{20-Dolgov:2002wy}.

Although neutrino decoupling is not described by a unique $T_{\rm
dec}$, it can be approximated as an instantaneous process.  The
standard picture of {\it instantaneous neutrino decoupling} is very
simple (see e.g.\ \cite{20-kt,20-dodelson}) and
reasonably accurate.  In this approximation, the spectrum in eq.\
(\ref{20-FD}) is preserved after decoupling, because both neutrino momenta
and temperature redshift identically with the expansion of the
Universe. In other words, the number density of non-interacting
neutrinos remains constant in a comoving volume since decoupling.
We will see later that neutrinos cannot possess masses
much larger than $1$ eV, so they were ultra-relativistic at
decoupling. This is the reason why the momentum distribution in eq.\
(\ref{20-FD}) does not depend on the neutrino masses, even after
decoupling, i.e.\ there is no neutrino energy in the exponential of
$f_{\rm eq}(p)$.

When calculating quantities related to relic neutrinos, one must
consider the various possible degrees of freedom per flavour. If
neutrinos are massless or Majorana particles, there are two degrees of
freedom for each flavour, one for neutrinos (one negative helicity
state) and one for antineutrinos (one positive helicity state).
Instead, for Dirac neutrinos there are in principle twice more degrees
of freedom, corresponding to the two helicity states. However, the
extra degrees of freedom should be included in the computation only if
they are populated and brought into equilibrium before 
neutrinos decouple. In practice, the Dirac neutrinos with the
``wrong-helicity'' states do not interact with the plasma at MeV
temperatures and have a vanishingly small density
with respect to the usual left-handed neutrinos (unless neutrinos have
masses close to the keV range, as explained in section 6.4 of
\cite{20-Dolgov:2002wy}, but this possibility is excluded). Thus the relic density of active neutrinos does not
depend on their nature, either Dirac or Majorana particles.

Shortly after neutrino decoupling the temperature drops below the
electron mass, favouring $e^{\pm}$ annihilations into
photons. If one assumes that this entropy transfer did not affect the
neutrinos because they were already completely decoupled, it is easy
to calculate the change in the photon temperature before any $e^{\pm}$
annihilation and after the $e^{\pm}$ pairs disappear by
assuming entropy conservation of the electromagnetic plasma. The
result, $T^{\rm after}_\gamma/T^{\rm before}_\gamma=(11/4)^{1/3}\simeq 1.40102$,
is also the ratio between the temperatures of relic photons and
neutrinos $T_\gamma/T_\nu$. During the process of $e^{\pm}$ annihilations  $T_\gamma$
decreases with the expansion less
than the inverse of the scale factor $a$. Instead the temperature of
the decoupled neutrinos always falls as $1/a$.

It turns out that the processes of neutrino
decoupling and $e^{\pm}$ annihilations are sufficiently close in time
so that some relic interactions between $e^{\pm}$ and neutrinos exist.
These relic processes are more efficient for larger neutrino energies,
leading to non-thermal distortions in the neutrino spectra at the percent level and a slightly smaller increase of the comoving photon
temperature, as noted in a series of works listed in \cite{20-NuCosmo,20-Dolgov:2002wy}.
%A proper calculation of the process
%of non-instantaneous neutrino decoupling demands solving the
%momentum-dependent Boltzmann equations for the neutrino spectra, a set
%of integro-differential kinetic equations that are difficult to solve
%numerically. This problem was considered in \cite{20-Mangano:2005cc}
%including the effect of flavour neutrino oscillations on
%the neutrino decoupling process. One finds an increase in the neutrino
%energy densities with respect to the instantaneous decoupling
%approximation ($0.73\%$ and $0.52\%$ for $\nu_e$'s and
%$\nu_{\mu,\tau}$'s, respectively) and a value of the comoving photon
%temperature after $e^\pm$ annihilations which is a factor $1.3978$
%larger, instead of $1.40102$. 
These changes modify the contribution of
relativistic relic neutrinos to the total energy density which is
taken into account using $N_{\rm eff} \simeq 3.046$ \cite{20-Mangano:2005cc}, as defined later
in eq.\ (\ref{20-neff}). In practice, these distortions only have small consequences on the
evolution of cosmological perturbations, and for many purposes they
can be safely neglected.

Any quantity related to relic neutrinos can be calculated after
decoupling with the spectrum in eq.\ (\ref{20-FD}) and $T_\nu$. For
instance, the number density per flavour is %fixed by the temperature,
\begin{equation}
n_{\nu} = \frac{3}{11}\;n_\gamma =
\frac{6\zeta(3)}{11\pi^2}\;T_\gamma^3~,
\label{20-nunumber}
\end{equation}
which leads to a present value of $113$ neutrinos and antineutrinos of
each flavour per cm$^{3}$. Instead, the energy density for massive
neutrinos should in principle be calculated numerically, with two
well-defined analytical limits,
\begin{eqnarray}
\rho_\nu (m_\nu \ll T_\nu)  =  
\frac{7\pi^2}{120}
\left(\frac{4}{11}\right)^{4/3}\;T_\gamma^4~,
%\label{20-rhomassless}
\qquad
\rho_\nu (m_\nu \gg T_\nu)  =  m_\nu n_\nu~.
\label{20-nurho}
\end{eqnarray}
%

%\subsection{Background evolution}
%\label{20-subsec:background}

Let us now discuss the evolution of the CNB after decoupling in the
expanding Universe, which is described by the
Friedmann-Robertson-Walker metric \cite{20-dodelson}
\begin{equation}
ds^2 = dt^2 - a(t)^2\,\delta_{ij} dx^i dx^j ~,
\label{20-Friedmann_metric}
\end{equation}
where we assumed negligible spatial curvature. Here $a(t)$ is the
scale factor usually normalized to unity now ($a(t_0)=1$) and related
to the redshift $z$ as $a=1/(1+z)$.  General relativity tells us the
relation between the metric and the matter and energy %in the Universe
via the Einstein equations, whose time-time component is the Friedmann
equation
\begin{equation}
\left(\frac{\dot{a}}{a}\right )^2=H^2 = \frac{8 \pi G}{3} 
\rho= H_0^2\frac{\rho}{\rho_{\rm c}^0}~,
\end{equation}
that gives the Hubble rate in terms of the total energy density
$\rho$.  At any time, the critical density $\rho_{\rm c}$ is defined
as $\rho_{\rm c}=3H^2/8\pi G$, with a current value
$\rho_{\rm c}^0 = 1.8788 \times 10^{-29}\,h^2 ~\mathrm{g~cm}^{-3}$,
where $h\equiv H_0/(100~{\rm km\, s}^{-1} \,{\rm Mpc}^{-1})$.
The different contributions to the total energy density are
$\rho=
\rho_{\gamma} + \rho_{\rm cdm} + \rho_{\rm b} 
+  \rho_{\nu} + \rho_{\Lambda}$,
and the evolution of each component is given by the energy
conservation law in an expanding Universe $\dot{\rho}=-3H(\rho+p)$,
where $p$ is the pressure. Thus the homogeneous density of photons
$\rho_{\gamma}$ scales like $a^{-4}$, that of non-relativistic matter
($\rho_{\rm cdm}$ for cold dark matter and $\rho_{\rm b}$ for baryons)
like $a^{-3}$, and the cosmological constant density $\rho_{\Lambda}$
is of course time-independent. Instead, the energy density of 
neutrinos contributes to the radiation density at early times but they
behave as matter after the non-relativistic transition. 

\begin{figure}[t]
\begin{center}
%\vspace{-2.5cm}
%\hspace{-1.5cm}
\includegraphics[width=.5\textwidth]{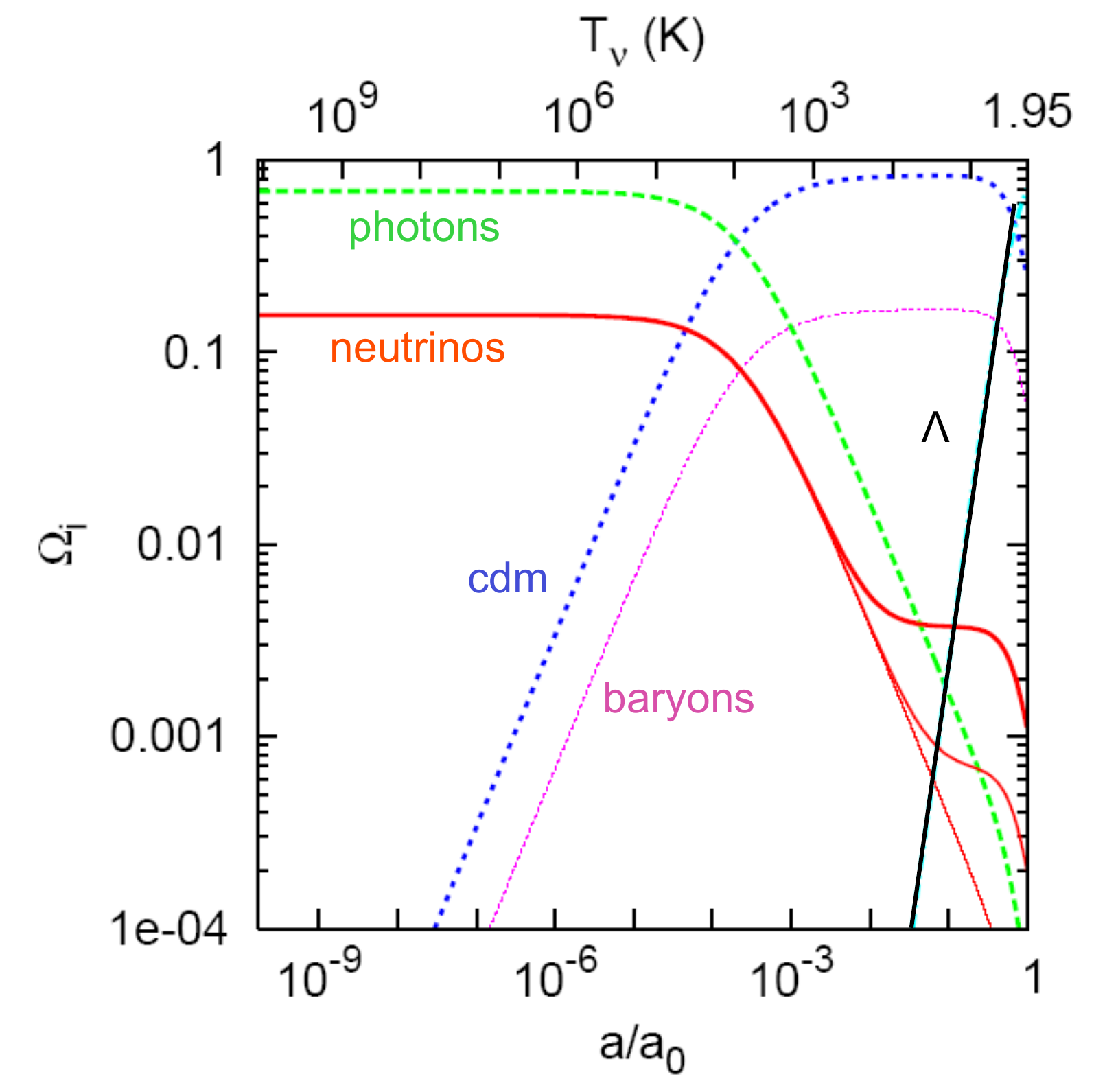}
\caption{\label{20-fig:rho_i} Evolution of the background energy
densities in terms of the fractions $\Omega_i$,
from $T_{\nu}=1$ MeV until now, for each component of a flat
Universe with $h=0.7$ and current density fractions
$\Omega_{\Lambda}=0.70$, $\Omega_{\rm b}=0.05$
and $\Omega_{\rm cdm}=1-\Omega_{\Lambda}-\Omega_{\rm b}
-\Omega_{\nu}$. The three neutrino masses are $m_1=0$, $m_2 = 0.009$
eV and $m_3 = 0.05$ eV.}
\end{center}
\end{figure}

The evolution of all densities is depicted in 
Fig.\ \ref{20-fig:rho_i}, starting at MeV temperatures until now. The
density fractions $\Omega_{i}\equiv\rho
_{i}/\rho_{\rm c}$ are shown in this figure, where it is easy to
see which of the Universe components dominantes, fixing its expansion
rate: first radiation in the form of photons and neutrinos (Radiation
Domination), then matter which can be CDM, baryons and massive
neutrinos at late times (Matter Domination) and finally the
cosmological constant density takes over at low $z$ (typically $z
< 0.5$).

Massive neutrinos are the only known particles that present a late transition
from radiation to matter, when their density is clearly enhanced
(upper solid lines in Fig.\ \ref{20-fig:rho_i}). Obviously the
contribution of massive neutrinos to the energy density in the
non-relativistic limit is a function of the mass (or the sum of all
masses for which $m_i \gg T_\nu$), and the present value $\Omega_\nu$
could be of order unity  for eV masses (see 
section \ref{20-sec:nuDM}).

\section{Extra radiation and the effective number of neutrinos}
\label{20-subsec:neff}

Together with photons, in the standard case neutrinos fix the
expansion rate while the Universe is
dominated by radiation. Their contribution to the total radiation
content can be parametrized in terms of the effective number of
neutrinos $N_{\rm eff}$, defined as %through the relation
\begin{equation}
\rho_{\rm r} = \rho_\gamma + \rho_\nu = 
\left[ 1 + \frac{7}{8} \left( \frac{4}{11}
\right)^{4/3} \, N_{\rm eff} \right] \, \rho_\gamma \,\,,
\label{20-neff}
\end{equation}
where we have normalized $\rho_{\rm r}$ to the photon energy density because its
value today is known from the measurement of the CMB temperature. This
equation is valid when neutrino decoupling is complete and holds as
long as all neutrinos are relativistic. 

We know that the number of light neutrinos sensitive to weak
interactions (flavour or active neutrinos) equals three from the
analysis of the invisible $Z$-boson width at LEP, 
$N_\nu=2.9840 \pm
0.0082$ \cite{Beringer:1900zz}, and we saw in a previous section from the
analysis of neutrino decoupling that they contribute as $N_{\rm
eff}\simeq 3.046$. Any departure of $N_{\rm eff}$ from this last value
would be due to non-standard neutrino features or to the contribution
of other relativistic relics. For instance, the energy density of a
hypothetical scalar particle $\phi$ in equilibrium with the same
temperature as neutrinos would be $\rho_\phi=(\pi/30)\,T_\nu^4$,
leading to a departure of $N_{\rm eff}$ from the standard value of
$4/7$.  A detailed discussion of cosmological scenarios where $N_{\rm
eff}$ is not fixed to three can be found in \cite{20-NuCosmo,20-Dolgov:2002wy,20-Sarkar:1995dd}. 

Relativistic particles, such as neutrinos, fix the expansion rate during Big Bang Nucleosynthesis 
(BBN)\footnote{In addition, BBN is the last cosmological epoch sensitive to neutrino flavour,
 because electron neutrinos and antineutrinos play a direct role in the weak processes.} which
 in turn fixes the produced abundances of light elements, and in particular
that of $^4$He. Thus the value of $N_{\rm eff}$ is constrained at
BBN from the comparison of theoretical predictions and
experimental data on the primordial abundances of light elements \cite{20-Steigman,20-Iocco:2008va}.  
This is why BBN gave the first allowed range of the number of neutrino
species before accelerators.  In a recent analysis of the BBN constraints on $N_{\rm eff}$ \cite{20-Mangano:2011ar} (see the references
therein for other analyses), the authors discuss a new and more conservative approach, motivated by growing concerns on the reliability of 
astrophysical determinations of primordial $^4$He. 
According to \cite{20-Mangano:2011ar}, BBN limits the extra radiation 
to $\Delta N_{\rm eff}\leq1$ at 95\% C.L.

In addition, a value of $N_{\rm eff}$ different from the standard one
would affect the CMB observables, as will be explained in section \ref{neff}. 
We will see that the Planck satellite measurement is in very good agreement with both the standard prediction
of $N_{\rm eff}\simeq 3.046$ and BBN results, in spite of a marginal preference 
for extra relativistic degrees of freedom (exacerbated if astrophysical
measurements of $H_0$ are included).

%would modify the transition epoch from a radiation-dominated to a
%matter-dominated Universe, which has  consequences on some
%cosmological observables such as the power spectrum of CMB
%anisotropies, leading to independent bounds on the radiation content.
%Until 2013, many analyses of late cosmological observables
%seemed to favour $N_{\rm eff}>3$, with best-fit values of
%order $4.3-4.4$, although with large errorbars of order $1.5$ at 95\% C.L.,
%as shown for instance in \cite{20-Komatsu:2010fb}
%(see also \cite{20-Dunkley:2010ge,20-Keisler:2011aw,20-Hamann:2011hu}).
%
%The recent CMB measurements by the Planck satellite 
%have pinned down the radiation content of the Universe \cite{Ade:2013zuv}: the allowed range is
%$N_{\rm eff}=3.30^{+0.54}_{-0.51}$
%(95\% C.L.; Planck, WMAP polarization, high-L CMB and BAO data, as defined later).
%This range is in very good agreement with both the standard prediction
%of $N_{\rm eff}\simeq 3.046$ and BBN results.
%There remains a marginal preference 
%for extra relativistic degrees of freedom (exacerbated if astrophysical
%measurements of $H_0$ are included), but this leaves
%small room for the presence of other particles such as 
%light sterile neutrinos, that could be populated via oscillations in the early Universe 
%\cite{20-Abazajian:2012ys,20-Hannestad:2012ky}.

\section{Massive neutrinos as Dark Matter}
\label{20-sec:nuDM}

Nowadays the existence of Dark Matter (DM), the dominant non-baryonic
component of the matter density in the Universe, is well established.
A priori, massive neutrinos are excellent DM candidates, in particular
because we are certain that they exist, in contrast with other
candidate particles. Their energy density in units of the
critical value is
\begin{equation}
\Omega_{\nu} = \frac{\rho_\nu}{\rho^0_{\rm c}} =
\frac{\sum_i m_i}{93.14\,h^2~{\rm eV}}~.
\label{20-omeganu}
\end{equation}
Here $\sum_i m_i$ includes all masses of the neutrino states which are
non-relativistic today. It is also useful to define the neutrino
density fraction with respect to the matter density
\begin{equation}
f_{\nu} \equiv \frac{\rho_{\nu}}{(\rho_{\rm cdm}+\rho_{\rm b}
+\rho_{\nu})} =
\frac{\Omega_{\nu}}{\Omega_{\rm m}}
\end{equation}

In order to find the contribution of relic neutrinos 
to the present values of $\Omega_{\nu}$ or $f_{\nu}$, we
should consider which neutrino masses are allowed by non-cosmological
data. Oscillation experiments measure $\Delta m^2_{21} = m^2_2 - m^2_1$ and $\Delta m^2_{31}
= m^2_3 - m^2_1$, the relevant differences of squared
neutrino masses for solar and atmospheric
neutrinos, respectively. 
As a reference, we take the
$3\sigma$ ranges of mixing parameters from 
\cite{20-Tortola:2012te} (see also \cite{20-Fogli:2012ua,20-GonzalezGarcia:2012sz}), 
\begin{eqnarray}
\fl 
s_{12}^2  =  0.32\pm{0.05}
\quad
s_{23}^2  \in [0.36,0.68]\, ([0.37,0.67])
\quad s_{13}^2  =  0.0246_{-0.0076}^{+0.0084}\, (0.025 \pm 0.008) \nonumber \\
\fl
\Delta m^2_{21} (10^{-5}~{\rm eV}^2)= 7.62_{-0.50}^{+0.58} \qquad
\Delta m^2_{31} (10^{-3}~{\rm eV}^2) =2.55_{-0.24}^{+0.19} \, (-2.43_{-0.22}^{+0.21})
\label{20-oscpardef}
\end{eqnarray}
Here $s^2_{ij}=\sin^2 \theta_{ij}$, where $\theta_{ij}$ ($ij=12, 23$ or
$13$) are the three mixing angles.  Unfortunately oscillation
experiments are insensitive to the absolute scale of neutrino masses,
because the knowledge of $\Delta m^2_{21}>0$ and $|\Delta m^2_{31}|$
leads to the two possible schemes shown in fig.\ 1 of
\cite{20-Lesgourgues:2006nd}, but leaves one neutrino mass unconstrained. These
two schemes are known as normal (NH, $\Delta m^2_{31}>0$) and inverted (IH, $\Delta m^2_{31}<0$) mass hierarchies.
In the above equation the values in parentheses 
correspond to the IH, otherwise the allowed regions are the same for both hierarchies. For small values of the lightest neutrino mass $m_0$,
i.e.\ $m_1$ ($m_3$) for NH (IH), the mass states follow a hierarchical
scenario, while for masses much larger than the differences all
neutrinos share in practice the same mass (degenerate region).
In general, the relation between the individual masses
and the total neutrino mass can be found numerically.%, as shown in Fig.\ \ref{20-fig:numasses}.
\begin{figure}[t]
\begin{center}
%\vspace{-2cm}
%\hspace{-1.5cm}
%\includegraphics[width=.48\textwidth]{lesgourgues/figure3a.pdf}\\
\hspace{-0.5cm}
\includegraphics[width=.52\textwidth]{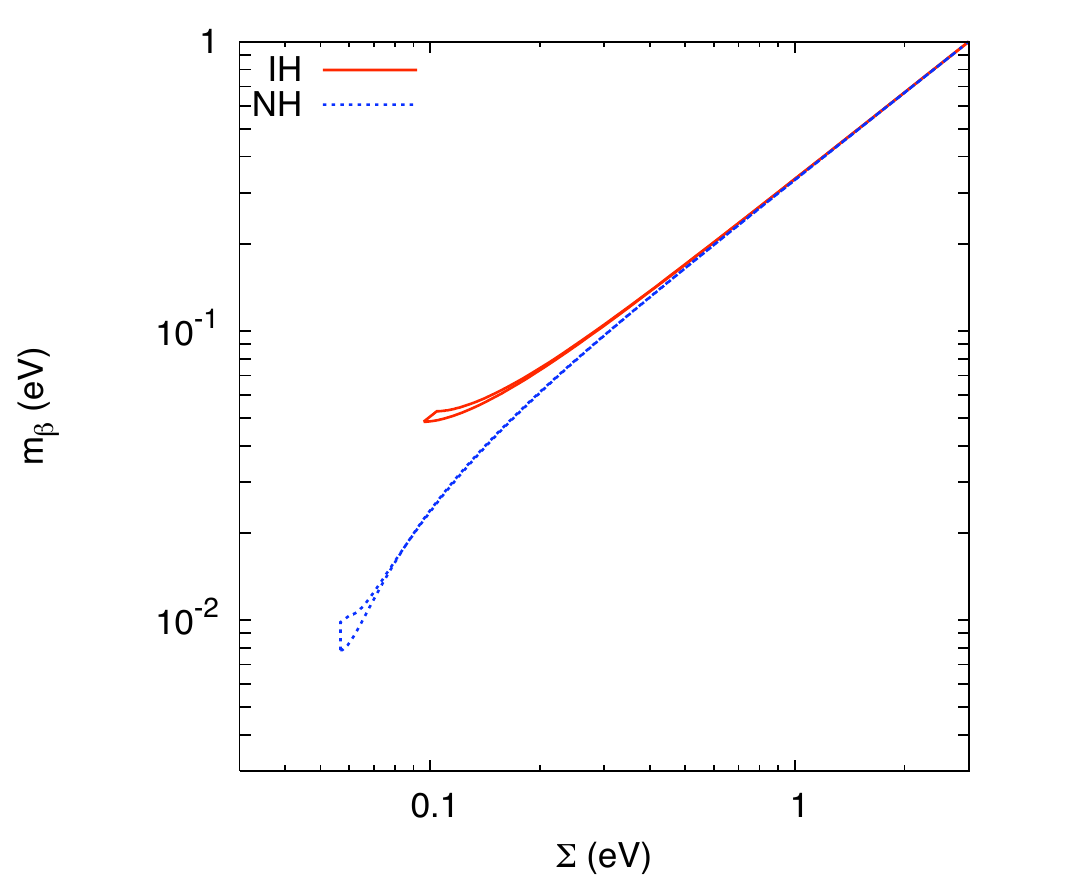}
\hspace{-0.5cm}
\includegraphics[width=.52\textwidth]{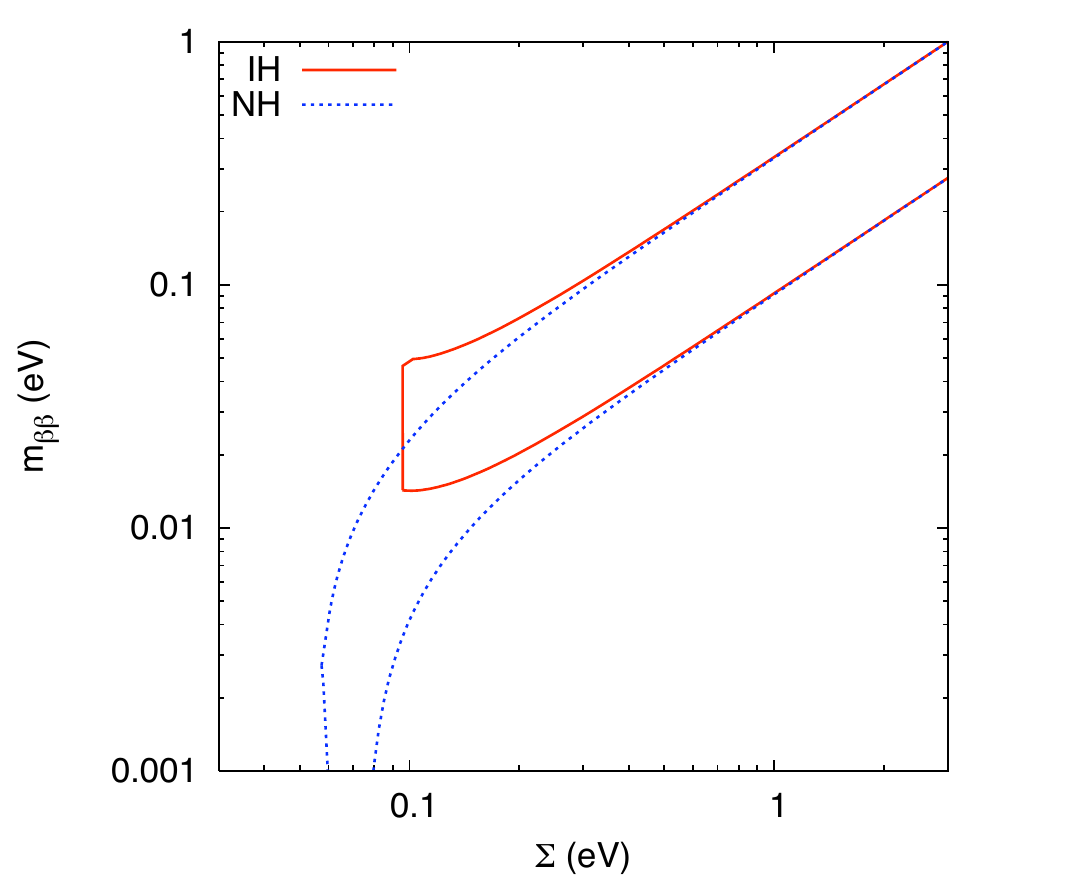}
\caption{\label{20-fig:numasses} 
Allowed regions by oscillation data at the 3$\sigma$ level (eq.\ (\ref{20-oscpardef}))
of the three main observables sensitive to the absolute scale of neutrino masses.  We show the regions
in the planes $m_{\beta}-\Sigma$ and $m_{\beta\beta}-\Sigma$, where $\Sigma$ is the sum of neutrino masses.
Blue dotted (red solid) regions correspond to normal (inverted) hierarchy.}
\end{center}
\end{figure}

There are two types of laboratory experiments searching for the
absolute scale of neutrino masses, a crucial piece of information for
constructing models of neutrino masses and mixings.
The neutrinoless double beta decay $(Z,A) \to (Z+2,A)+2e^-$ (in short
$0\nu2\beta$) is a rare nuclear processes where lepton number is
violated and whose observation would mean that neutrinos are Majorana
particles. If the $0\nu2\beta$ process is mediated by a light
neutrino, the results from neutrinoless double beta decay experiments
are converted into an upper bound or a measurement of the effective
mass % $m_{\beta\beta}$
\begin{equation}
m_{\beta\beta} %= \left |\sum_i U_{ei}^2\,  m_i\right |
= |c_{12}^2c_{13}^2\, m_1+s_{12}^2c_{13}^2\, m_2\, e^{i\phi_2}
+s_{13}^2\, m_3\, e^{i\phi_3}|~,
\label{20-beta2beta}
\end{equation}
where $\phi_{1,2}$ are two Majorana phases that appear in
lepton-number-violating processes. An important issue for $0\nu2\beta$
results is related to the uncertainties on the corresponding nuclear
matrix elements. For more details
and the current experimental results, see \cite{20-Giuliani}.

Beta decay experiments, which involve only the kinematics of
electrons, are in principle the best strategy for measuring directly
the neutrino mass \cite{20-Drexlin}.  The current limits from tritium
beta decay apply only to the range of degenerate neutrino masses, so
that $m_\beta \simeq m_0$, where
\begin{equation}
m_\beta %= \left (\sum_i |U_{ei}|^2\, m_i^2\right )^{1/2}
= (c_{12}^2c_{13}^2\, m_1^2+s_{12}^2c_{13}^2\, m_2^2
+s_{13}^2\, m_3^2)^{1/2},
\label{20-beta}
\end{equation}
is the relevant parameter for beta decay experiments. The bound at
$95\%$ CL is $m_0<2.05-2.3$ eV from the Troitsk and Mainz experiments,
respectively. This value is expected to be improved by the KATRIN
project to reach a discovery potential for $0.3-0.35$ eV masses (or a
sensitivity of $0.2$ eV at $90\%$ CL). Taking into account this upper
bound and the minimal values of the total neutrino mass in the normal
(inverted) hierarchy, the sum of neutrino masses $M_\nu \equiv \sum_i m_i$ is restricted to the
approximate range
\begin{equation}
0.06\, (0.1)~{\rm eV} \lesssim M_\nu \lesssim 6~{\rm eV}
\label{20-sum_range}
\end{equation}

As we discuss in the next sections, cosmology is at first order
sensitive to the total neutrino mass if all states have the same
number density, providing information on $m_0$ but blind to neutrino
mixing angles or possible CP violating phases. Thus cosmological
results are complementary to terrestrial experiments. The interested
reader can find the allowed regions in the parameter space defined by
any pair of parameters $(M_\nu,m_{\beta\beta},m_\beta)$ in
\cite{20-Fogli:2005cq,20-Fogli:2008ig,20-GonzalezGarcia:2010un}.
The two cases involving $M_\nu$ are shown in Fig.\ \ref{20-fig:numasses}.

Now we can find the possible present values of $\Omega_\nu$ in
agreement  the approximate bounds of eq.\ (\ref{20-sum_range}).
Note that even if the three neutrinos are non-degenerate in mass, eq.\
(\ref{20-omeganu}) can be safely applied, because we know from neutrino
oscillation data that at least two of the neutrino states are
non-relativistic today, because both $|\Delta m^2_{31}|^{1/2}\simeq 0.05
$ eV and $(\Delta m^2_{21})^{1/2}\simeq 0.009$ eV are larger than the
temperature $T_\nu \simeq 1.96$ K $\simeq 1.7\times 10^ {-4}$ eV. If
the third neutrino state is very light and still relativistic, its
relative contribution to $\Omega_{\nu}$ is negligible and eq.\
(\ref{20-omeganu}) remains an excellent approximation of the total
density. One finds that $\Omega_\nu$ is restricted to the approximate
range
\begin{equation}
0.0013\, (0.0022) \lesssim \Omega_{\nu} \lesssim 0.13
\label{20-omega_nu_limits}
\end{equation}
where we already included that $h\approx 0.7$. This applies only to
the standard case of three light active neutrinos, while in general a
cosmological upper bound on $\Omega_{\nu}$ has been used since the
1970s to constrain the possible values of neutrino masses. For
instance, if we demand that neutrinos should not be heavy enough to
overclose the Universe ($\Omega_{\nu}<1$), we obtain an upper bound
$M_\nu \lesssim 45$ eV (again fixing $h=0.7$). Moreover, from
present analyses of cosmological data we know that the approximate
contribution of matter is $\Omega_{\rm m} \simeq 0.3$, the neutrino
masses should obey the stronger bound $M_\nu \lesssim 15$ eV. Thus
with this simple argument one obtains a bound which is roughly
only a factor 2 worse than the bound from tritium beta decay, but of
course with the caveats that apply to any cosmological analysis.  In
the three-neutrino case, these bounds should be understood in terms of
$m_0=M_\nu/3$.

Dark matter particles with a large velocity dispersion such as that of
neutrinos are called hot dark matter (HDM). The role of neutrinos as
HDM particles has been widely discussed since the 1970s (see e.g.\ the historical review \cite{20-Primack:2001ib}).
It was realized in the mid-1980s that HDM affects the evolution of
cosmological perturbations in a particular way: it erases the density
contrasts on wavelengths smaller than a mass-dependent free-streaming
scale. In a universe dominated by HDM, this suppression 
contradicts various observations. For instance, large objects
such as superclusters of galaxies form first, while smaller structures
like clusters and galaxies form via a fragmentation process. This
top-down scenario is at odds with the fact that galaxies seem older
than clusters.

Given the failure of HDM-dominated scenarios, the attention then
turned to cold dark matter (CDM) candidates, i.e.\ particles which
were non-relativistic at the epoch when the universe became
matter-dominated, which provided a better agreement with
observations. Still in the mid-1990s it appeared that a small mixture
of HDM in a universe dominated by CDM fitted better the observational
data on density fluctuations at small scales than a pure CDM
model. However, within the presently favoured $\Lambda$CDM model
dominated at late times by a cosmological constant (or some form of
dark energy) there is no need for a significant contribution of
HDM. Instead, one can use the available cosmological data to find how
large the neutrino contribution can be, as we will see later.

Before concluding this section, we would like to mention the case of a
sterile neutrino with a mass of the order of a few keV's and a very
small mixing with the flavour neutrinos. Such ``heavy'' neutrinos
could be produced by active-sterile oscillations but not fully
thermalized, so that they could %play the role of dark matter and
replace the usual CDM component. But due to their large thermal
velocity (slightly smaller than that of active neutrinos), they would
behave as Warm Dark Matter and erase small-scale cosmological
structures.  Their mass can be bounded from below using Lyman-$\alpha$
forest data from quasar spectra, and from above using X-ray
observations. The viability of this scenario %is currently under
%careful examination (see e.g.\ \cite{20-Boyarsky:2009ix}).
was reviewed in \cite{20-Boyarsky:2009ix}.

\section{Effects of neutrino masses on cosmology}
\label{20-sec:effects_numass}

Here we  briefly describe the main cosmological
observables, and their sensitivity to neutrino masses.  A
more detailed discussion of the effects of massive neutrinos on the
evolution of cosmological perturbations can be found in 
sections 5.3.3 and 6.1.4 of  \cite{20-NuCosmo}.

\subsection{Brief description of cosmological observables}

Although there exist many different types of cosmological
measurements, here we will restrict the discussion to those that are
at present the more important for obtaining an upper bound or
eventually a measurement of neutrino masses.

First of all, we have the CMB temperature anisotropy power spectrum, defined as the angular two-point correlation
function of CMB maps $\delta T/\bar{T}(\hat{n})$ ($\hat{n}$ being a
direction in the sky). This function is usually expanded in Legendre
multipoles
\begin{equation}
\left\langle {\delta T\over\bar{T}}(\hat{n}) 
{\delta T \over \bar{T}} (\hat{n}') \right\rangle
= \sum_{l=0}^{\infty} {(2l+1)\over4\pi} \,C_l\, P_l(\hat{n}\cdot\hat{n}')~,
\end{equation}
where $P_l(x)$ are the Legendre polynomials. For Gaussian
fluctuations, all the information is encoded in the multipoles $C_l$
which probe correlations on angular scales $\theta=\pi/l$.  
%We have
%seen that each neutrino state can only have a mass of the order of
%$1$ eV, so that the transition of relic neutrinos to the
%non-relativistic regime is expected to take place after the time of
%recombination between electrons and nucleons, i.e.\ after photon
%decoupling. Because the shape of the CMB spectrum is related mainly to
%the physical evolution {\it before} recombination, it will only be
%marginally affected by the neutrino mass, except through a modified background evolution and some secondary anisotropy corrections. 
There exists
interesting complementary information to the temperature power
spectrum if the CMB polarization is measured, and currently we have
some less precise data on the temperature $\times$ E-polarization (TE)
correlation function and the E-polarization self-correlation spectrum
(EE).

The current Large Scale Structure (LSS) of the Universe is probed by
the matter power spectrum at a given time or redhsift $z$. It is defined as the two-point correlation function
of non-relativistic matter fluctuations in Fourier space
\begin{equation}
P(k,z)=\langle | \delta_{\rm m}(k,z) |^2 \rangle~,
\end{equation}
where $\delta_{\rm m}=\delta \rho_{\rm m}/\bar{\rho}_{\rm m}$.
Usually $P(k)$ refers to the matter power spectrum evaluated today (at
$z=0$).  In the case of several fluids (e.g.\ CDM, baryons and
non-relativistic neutrinos), the total matter perturbation can be
expanded as
$\delta_{\rm m}= \sum_i \, \bar{\rho}_i \, \delta_i/\sum_i \, \bar{\rho}_i$.
Because the energy density is related to the mass density of
non-relativistic matter through $E=mc^2$, $\delta_{\rm m}$ represents
indifferently the energy or mass power spectrum. The shape of the
matter power spectrum is affected in a scale-dependent way by the
free-streaming caused by small neutrino masses of ${\cal O}$(eV) and
thus it is the key observable for constraining $m_\nu$ with
cosmological methods.

\subsection{Neutrino free-streaming}

After thermal decoupling, relic neutrinos constitute a collisionless
fluid, where the individual particles free-stream with a
characteristic velocity that, in average, is the thermal velocity
$v_{\rm th}$. It is possible to define a horizon as the typical
distance on which particles travel between time $t_i$ and $t$.  When
the Universe was dominated by radiation or matter $t \gg t_i$, this
horizon is, as usual, asymptotically equal to $v_{\rm th}/H$, up to a
numerical factor of order one. Similar to the definition of the Jeans
length (see section 4.4 in \cite{20-Lesgourgues:2006nd}), we can define the
neutrino free-streaming wavenumber %and length as
%
%\begin{eqnarray}
\begin{equation}
%\!\!\!\!\!\! 
k_{FS}(t) = \left(\frac{4 \pi G \bar{\rho}(t) 
a^2(t)}{v_{\rm th}^2(t)}\right)^{1/2}. \\
%\qquad
%\lambda_{FS}(t) 
%= 2 \pi \frac{a(t)}{k_{FS}(t)}
%= 2 \pi \sqrt{2 \over 3} \frac{v_{\rm th}(t)}{H(t)}~.
%\end{eqnarray}
\end{equation}
As long as neutrinos are relativistic, they travel at the speed of
light and their free-streaming length is simply equal to the Hubble
radius. When they become non-relativistic, their thermal velocity
decays like
\begin{equation}
v_{\rm th}\equiv\frac{\langle p \rangle}{m}
\simeq\frac{3.15 T_{\nu}}{m}=\frac{3.15 T_{\nu}^0}{m}
\left( \frac{a_0}{a} \right) 
\simeq 158 (1+z) \left( \frac{1 \, \mathrm{eV}}{m} \right)
{\rm km}\,{\rm s}^{-1}~,
\end{equation}
where we used for the present neutrino temperature $T_{\nu}^0 \simeq
(4/11)^{1/3} T_{\gamma}^0$ and $T_{\gamma}^0 \simeq 2.726$ K. This
gives for the free-streaming 
%wavelength and 
wavenumber during matter or ${\Lambda}$ domination
%
%\begin{eqnarray}
\begin{equation}
k_{FS}(t) = 0.8
\frac{\sqrt{\Omega_{\Lambda}+ \Omega_{m} (1+z)^3}}{(1+z)^2}
\left( \frac{m}{1 \, \mathrm{eV}} \right) h\,\mathrm{Mpc}^{-1}, 
\end{equation}
%\\
%\lambda_{FS}(t) 
%&=& 
%8 
%\frac{1+z}{\sqrt{\Omega_{\Lambda}+ \Omega_{m} (1+z)^3}}
%\left( \frac{1 \, \mathrm{eV}}{m} \right) h^{-1}\mathrm{Mpc}~,
%\end{eqnarray}
%
where $\Omega_{\Lambda}$ and $\Omega_{m}$ are the cosmological
constant and matter density fractions, respectively, evaluated today.
After the non-relativistic transition and during matter
domination, the free-streaming length continues to increase, but only
like $(aH)^{-1}\propto t^{1/3}$, i.e.\ more slowly than the scale
factor $a \propto t^{2/3}$.  Therefore, the comoving free-streaming
length $\lambda_{FS} / a$ actually decreases like $(a^2 H)^{-1}
\propto t^{-1/3}$. As a consequence, for neutrinos becoming
non-relativistic during matter domination, the comoving free-streaming
wavenumber passes through a minimum $k_{\rm nr}$ at the time of the
transition, i.e.\ when $m = \langle p \rangle = 3.15 T_{\nu}$ and
$a_0/a=(1+z)= 2.0\times 10^3 (m/ 1 \, \mathrm{eV})$. This minimum
value is found to be
\begin{equation}
k_{\rm nr}
\simeq 0.018 \,\, \Omega_{\rm m}^{1/2} 
\left( \frac{m}{1 \, \mathrm{eV}} \right)^{1/2}
h\,\mathrm{Mpc}^{-1}~.
\label{20-knr}
\end{equation}
The physical effect of free-streaming is to damp small-scale neutrino
density fluctuations: neutrinos cannot be confined into (or kept
outside of) regions smaller than the free-streaming length, because their velocity is greater than the escape velocity from gravitational potential wells on those scales.
Instead, on scales much larger than the free-streaming scale, the
neutrino velocity can be effectively considered as vanishing, and after
the non-relativistic transition the neutrino perturbations behave like
CDM perturbations. In particular, modes with $k<k_{\rm nr}$ are never
affected by free-streaming and evolve like being in a pure $\Lambda$CDM
model.

\subsection{Impact of massive neutrinos on the matter power spectrum}
\label{20-subsec:impact_pk}

The small initial cosmological perturbations evolve within the linear regime at early times. During matter domination, the smallest cosmological scales start evolving non-linearily, leading to the formation of the structures we see today. In the recent universe, the largest observable scales are still related to the linear evolution, while other scales can only be understood using non-linear N-body simulations. We will not review here all the details of this complicated evolution 
(see \cite{20-NuCosmo,20-Wong:2011ip,20-Lesgourgues:2006nd} and
references therein), but we will emphasize the main effects caused by
massive neutrinos on linear scales in the framework of the standard cosmological
scenario: a $\Lambda$ Mixed Dark Matter ($\Lambda$MDM) model, where
Mixed refers to the inclusion of some HDM component.

On large scales (i.e. on wave-numbers smaller than the value $k_{\rm nr}$ defined in the previous subsection), neutrino free-streaming can be ignored, and neutrino perturbations are indistinguishable from CDM perturbations. On those scales, the matter power spectrum $P(k,z)$ can be shown to depend only on the matter density fraction today (including neutrinos), $\Omega_{\rm m}$, and on the primordial perturbation spectrum. If the neutrino mass is varied with $\Omega_{\rm m}$ fixed, the large -scale power spectrum remains invariant.

\begin{figure}[t]
\begin{center}
\vspace{-6.5cm}
\includegraphics[width=0.65\textwidth]{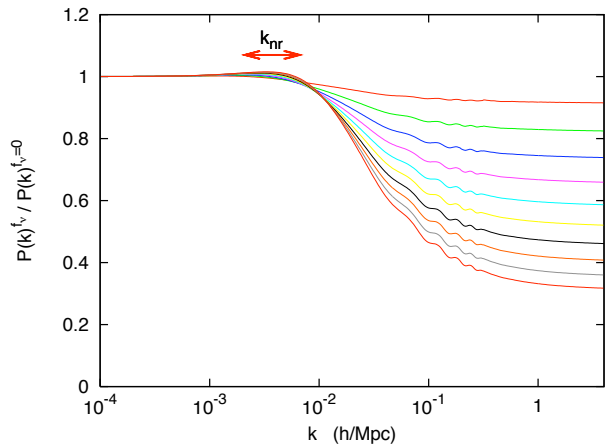}\\
\caption{\label{20-fig_tk} Ratio of the matter power spectrum including
three degenerate massive neutrinos with density fraction $f_{\nu}$ to
that with three massless neutrinos.  The parameters $(\omega_{\rm m},
\, \Omega_{\Lambda})=(0.147,0.70)$\ are kept fixed, and from top to
bottom the curves correspond to $f_{\nu}=0.01, 0.02, 0.03,\ldots,0.10$.
The individual masses $m_{\nu}$ range from $0.046$ to $0.46$ eV,
and the scale $k_{\rm nr}$ from $2.1\times10^{-3}h\,$Mpc$^{-1}$ to
$6.7\times10^{-3}h\,$Mpc$^{-1}$ as shown on the top of the
figure. {}From \cite{20-Lesgourgues:2006nd}.}
%$k_{\rm eq}$ is approximately equal to $1.5\times10^{-2}h\,$Mpc$^{-1}$.}
\end{center}
%\vspace{0.5cm}
\end{figure}

However, the small-scale matter power spectrum $P(k \geq k_{\rm nr})$ is reduced in presence of massive neutrinos for at least two reasons: by the absence of neutrino perturbations in the total matter power spectrum, and by a slower growth rate of CDM/baryon perturbations at late times. The third effect has the largest amplitude. At low redhsift $z \simeq 0$, the step-like suppression of $P(k)$ starts at $k \geq k_{\rm nr}$ and saturates at $k \sim 1 h/$Mpc with a constant amplitude $\Delta P(k)/P(k) \simeq -8 f_\nu$. This result was obtained by fitting numerical simulations \cite{20-Hu:1997mj}, but a more accurate approximation can be derived analytically 
\cite{20-NuCosmo,20-Lesgourgues:2006nd}. 
%As mentioned in the second item above, neutrino masses can have additional indirect effects through a change in the background evolution, depending on which cosmological parameters are kept fixed when $M_\nu$ varies.
%
%When fitting data, one can use
%analytical approximations to the full MDM or $\Lambda$MDM matter power
%spectrum, valid for arbitrary scales and redshifts, as listed in
%\cite{20-Lesgourgues:2006nd}. However, nowadays the analyses are performed
The full matter power spectrum can be calculated at any time and scale by Boltzmann codes such as 
{\sc camb} \cite{20-Lewis:1999bs} or {\sc class} \cite{20-Lesgourgues:2011rh},
that solve numerically the evolution of the cosmological
perturbations.  The step-like suppression of the matter power spectrum induced by various values of $f_\nu$ is shown in Fig.\ \ref{20-fig_tk}.

On small scales and at late times, matter density perturbations enter into the regime of non-linear clustering. 
%The value of $k_{\rm max}$ is not limited by observational sensitivities, but by the range in which we trust predictions from linear theory. Beyond the scale of non-linearity (of the order of $k_{\rm max} = 0.15 h\,$Mpc$^{-1}$ at $z=0$), the data is only useful provided that one is able to make accurate predictions not only for the non-linear power spectrum, but also for redshift-space distortions (coming from the fact that we observe the redshift of objects, not their distance away from us), and finally for the scale dependence of the light-to-mass bias (relating the  power spectrum of a given category of observed compact objects to the underlying total matter power spectrum). Spectacular progresses are being carried on these three fronts, thanks to better N-body simulations and analytical techniques. Including massive neutrinos in such calculations is particularly difficult, but successful attempts were presented e.g. in \cite{20-Brandbyge:2009ce,20-Bird:2011rb,20-Wagner:2012sw}. 
The neutrino mass impact on the non-linear matter power spectrum is now modeled with rather good precision, at least  within one decade above $k_{\rm max}$ in wave-number space \cite{20-Brandbyge:2009ce,20-Bird:2011rb,20-Wagner:2012sw}. It appears that the step-like suppression is enhanced by non-linear effects up to roughly $\Delta P(k)/P(k) \simeq -10 f_\nu$ (at redshift zero and $k\sim 1\,h\,$Mpc$^{-1}$), and is reduced above this scale. Hence, non-linear corrections render the neutrino mass effect even more characteristic than in the linear theory, and may help to increase the sensitivity of future experiments 
(see also \cite{Villaescusa-Navarro:2013pva,Castorina:2013wga,Costanzi:2013bha} for the effect of neutrino masses on even smaller scales).

Until this point, we reduced the neutrino mass effect to that of $f_\nu$ or $M_\nu$. In principle, the mass splitting between the three different families for a common total mass is visible in $P(k)$. The time at which each species becomes non-relativistic depends on individual masses $m_i$. Hence, both the scale of the step-like suppression induced by {\it each} neutrino and the amount of suppression in the small-scale power spectrum have a small dependence on individual masses. The differences between the power spectrum of various models with the same total mass and different mass splittings was computed numerically in \cite{20-Lesgourgues:2004ps} for the linear spectrum, and \cite{20-Wagner:2012sw} for the non-linear spectrum. At the moment, it seems that even the most ambitious future surveys will not be able to distinguish these mass splitting effects with a reasonable significance \cite{20-Jimenez:2010ev,20-Pritchard:2008wy}.

\subsection{Impact of massive neutrinos on the CMB anisotropy spectrum}
\label{20-subsec:impact_cl}

For masses smaller than 0.6 eV,
neutrinos are still relativistic at the time of
photon decoupling, and their mass cannot impact the evolution of CMB perturbations. Therefore, the effect
of the mass can only appear at two levels: that of the background evolution, and that of secondary anisotropies (related to the behaviour of photon perturbations after decoupling: Integrated Sachs-Wolfe (ISW) effect, weak lensing by large scale structure, etc.)

Let us first review the background effects of massive neutrinos on  the CMB. Because the temperature and polarization spectrum shape is the result of several intricate effects, one cannot discuss the neutrino mass impact without specifying which other parameters are kept fixed. Neutrinos with a mass in the range from $10^{-3}$ eV to $1$ eV should be counted as radiation at the time of equality, and as non-relativistic matter today: the total non-relativistic density, parametrized by $\omega_{\rm m}=\Omega_{\rm m} h^2$, depends on the total neutrino mass $M_\nu = \sum_i m_i$. Hence, when $M_\nu$ is varied, there must be a variation either in the redshift of matter-to-radiation equality $z_{\rm eq}$, or in the matter density today $\omega_{\rm m}$.

This can potentially impact the CMB in three ways. A shift in the redshift of equality affects the position and amplitude of the peaks. A change in the non-relativistic matter density at late times can impact both the angular diameter distance to the last scattering surface $d_A(z_{\rm dec})$, controlling the overall position of CMB spectrum features in multipole space, and the slope of the low-$l$ tail of the CMB spectrum, due to the late Integrated Sachs-Wolfe (ISW) effect. Out of these three effects (changes in $z_{\rm eq}$, in $d_A$ and in the late ISW), only two can be cancelled by a simultaneous variation of the total neutrino mass and of other free parameters in the $\Lambda$MDM model. Hence, the CMB spectrum is sensitive to the background effect of the total neutrino mass. In practice however, the late ISW effect is difficult to measure due to cosmic variance and CMB data alone cannot provide a useful information on sub-eV neutrino masses. If one considers extensions of the $\Lambda$MDM, this becomes even more clear: by playing with the spatial curvature, one can neutralize all three effects simultaneously. But as soon as CMB data is used in combination with other background cosmology observations (constraining for instance the Hubble parameter, the cosmological constant value or the BAO scale), some bounds can be derived on $M_\nu$.

There exists another effect of massive neutrinos on the CMB at the level of secondary anisotropies: when neutrinos become non-relativistic, they  reduce the time variation  of the gravitational potential inside the Hubble radius. This affects the photon temperature through the early ISW effect, and leads to a depletion in the temperature spectrum of the order of $(\Delta C_l/C_l) \sim -(M_\nu/0.3 \, {\rm eV}) \%$  on multipoles $20<l<500$, with a dependence of the maximum depletion scale on individual masses $m_i$ \cite{20-NuCosmo,20-Hou:2012xq}. This effect is roughly thirty times smaller than the depletion in the small-scale matter power spectrum, $\Delta P(k)/P(k) \sim -(M_\nu/0.01 \, {\rm eV}) \%$.

\begin{figure}[t]
\begin{center}
\includegraphics[width=0.65\textwidth]{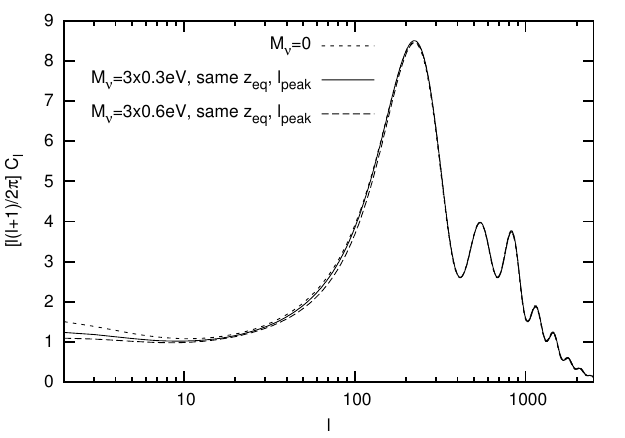}\\
\caption{\label{20-fig_cmb} CMB temperature spectrum with different neutrino masses. Some of the parameters of the $\Lambda$MDM model have been varied together with $M_\nu$ in order to keep fixed the redshift of equality and the angular diameter distance to last scattering.}
\end{center}
\end{figure}

We show in Figure~\ref{20-fig_cmb}  the effect on the CMB temperature spectrum of increasing the neutrino mass while keeping $z_{\rm eq}$ and $d_A$ fixed: the only observed differences are then for $2<l<50$ (late ISW effect due to neutrino background evolution) and for $50<l<200$ (early ISW effect due to neutrino perturbations).
We conclude that the CMB alone is not a very powerful tool for constraining sub-eV neutrino masses, and should be used in combination with homogeneous cosmology constraints and/or measurements of the LSS power spectrum, for instance from  galaxy clustering, galaxy lensing or CMB lensing.% (see section \ref{20-sec:future}).

\section{Current bounds on neutrino masses}
\label{20-sec:current}

In this section, we review the neutrino mass bounds that can be derived from current cosmological data. Note that the confidence limits in the next subsections are all based on the Bayesian inference method, and are given at the 95\% confidence level after marginalization over all free cosmological parameters. We refer the reader to section 5.1 of \cite{20-Lesgourgues:2006nd} for a detailed discussion on this statistical method.

\subsection{CMB anisotropy bounds}
\label{20-sec:present_CMB}

The best measurement of CMB temperature anisotropies comes from the first release of the Planck satellite \cite{Ade:2013ktc,Ade:2013zuv}, which significantly improved over the nine-year data of WMAP \cite{Hinshaw:2012aka} on large angular scales, and over ground-based or balloon-borne (ACT, SPT, \ldots) on small scales. Moreover, the Planck temperature data offers the advantage of covering all relevant angular scales with a single experiment, hence avoiding relative calibration issues between different data sets. The publication of polarisation data has been postponed to the next Planck release. In the meantime, Planck temperature data is usually analysed together with the $l \leq 23$ multipoles of WMAP E-type polarisation (referred later as WP), in order to remove degeneracies between the optical depth to reionisation and other parameters. Moreover, Planck temperature data is used only up to the multipole $l=2500$: for larger $l$'s, it is better to use higher-resolution experiments (ACT, SPT), limited to $l>2500$ to avoid any overlap, and called usually``highL''. Note that including information on temperature at $l>2500$ can be useful for separating the various foregrounds from the CMB signal, but it has a very minor impact on cosmological parameters, which are well constrained by the Planck+WP data alone~\cite{Ade:2013zuv}.

Assuming the minimal $\Lambda$CDM model with six free parameters, and promoting the total neutrino mass $M_\nu$ as a seventh free parameter, one gets $M_\nu < 0.66$~eV (95\%; Planck+WP+highL) \cite{Ade:2013zuv}. As explained in section \ref{20-subsec:impact_cl}, this bound comes mainly from two physical effects: the early-ISW-induced dip at intermediate $l$'s, and the lensing effect causing a smoothing of the power spectrum at higher $l$'s. To show that the latter effect is important, one can treat the amplitude of the lensing potential spectrum as a free parameter (while in reality, its value is fully predicted by the underlying cosmological model). This is equivalent to partially removing the piece of information coming from the lensing of the last scattering surface. Then, the mass bound degrades by 63\%.

Assuming an extended underlying cosmological model, the mass bound gets weaker. This is true especially if one allows for non-zero spatial curvature in the universe: the bounds then degrades to $M_\nu < 0.98$~eV (95\%; Planck+WP+highL). However, when the density of extra relativistic relics is promoted as a new free parameter, the bound on $M_\nu$ does not change significantly, showing the current CMB data is able to resolve the degeneracy between $M_\nu$ and $N_\mathrm{eff}$, that used to exist when using older data sets.

\subsection{Adding information on the cosmological expansion}

The measurement of CMB anisotropies can be complemented by some data on the cosmological expansion at low redshift ($z<2$). The recent expansion history can be inferred from the luminosity of type-Ia supernovae, from the angular scale of Baryon Acoustic Oscillations (BAO) reconstructed from the galaxy power spectrum at various redshifts, or from direct measurements of the current expansion rate $H_0$ through nearby cepheids and supernovae.

These measurements are very useful probes of neutrino masses, not directly but indirectly. Indeed, we have seen in section \ref{20-subsec:impact_cl} that if the neutrino mass is varied at the same time as parameters controlling the late cosmological evolution (like $\Omega_\Lambda$ or $\Omega_k$ in a non-flat universe), most characteristic quantites affecting the shape of the CMB spectrum can be kept constant, including the angular scale of the first acoustic peak. Then, the effect of $M_\nu$ reduces to an irrelevant late ISW effect, plus the early ISW and lensing effects discussed above. But if instead these parameters are measured directly and constrained independently, the degeneracy is removed, and varying $M_\nu$ will also change the scale of the first peak. In that way, the CMB is very sensitive to $M_\nu$, since the peak scale is the quantity best constrained by CMB observations.

The problem is that as long as one assumes a minimal $\Lambda$CDM model, the combination of CMB data with either BAO, $H_0$ or supernovae data reveals small tensions (roughly at the 2-$\sigma$ level). This could mean either that some data sets have slightly underestimated systematics, or that something is not correct in the assumed underlying cosmological model. The main tension is between Planck+WP and HST measurements of $H_0$, giving respectively $H_0=67.3\pm2.54$~km~s$^{-1}$Mpc$^{-1}$ (95\%; Planck+WP) and $H_0=73.8\pm4.8$~km~s$^{-1}$Mpc$^{-1}$ (95\%; HST) \cite{Riess:2011yx}. BAO data are more consistent with Planck than with HST and exacerbate the tension, giving $H_0=67.8\pm1.5$~km~s$^{-1}$Mpc$^{-1}$ (95\%; Planck+WP+BAO). Note that direct $H_0$ measurements probe the local value of the expansion parameter, while BAO and CMB data measure the expansion rate averaged over very large scales. However, in the standard cosmological model, the local and global $H_0$ are expected to differ by a tiny amount, so the tension cannot be explained in that way~\cite{Marra:2013rba}. Promoting the total neutrino mass as a seventh free parameter does {\it not} release the tension either, since for the $\Lambda$CDM$+M_\nu$ model one finds $H_0=67.7\pm1.8$ km~s$^{-1}$ Mpc$^{-1}$ (95\%; Planck+WP+BAO) \cite{Ade:2013zuv}. 

The neutrino mass bound also depends on which data is used. For CMB+BAO, one gets $M_\nu < 0.23$ eV (95\%; Planck+WP+highL+BAO), while for CMB+HST one gets a stronger bound $M_\nu < 0.18$ eV (95\%; Planck+WP+highL+BAO), and for CMB + supernovae the result is slightly looser, $M_\nu < 0.25$ eV (95\%; Planck+WP+highL+SNLS) ~\cite{Ade:2013zuv} (some of these numbers are not given in the Planck paper, but in the publicly released grid of bounds available at the ESA website\footnote{{\tt http://www.sciops.esa.int/wikiSI/planckpla/} in section 6.9.}). If one takes the conservative point of view that the BAO data is the less likely to be affected by systematics because it is a pure geometrical measurement of an angle in the sky, one should mainly remember the first of these bounds (0.23~eV). This is already a very spectacular result, only a factor four higher than the minimum allowed value $M_\nu \sim 0.06$ eV.
 
\subsection{Adding information on large scale structure}

Large scale structure can be probed by Planck itself in several ways, using lensing extraction, foregrounds or spectral distorsions. 

Among these probes, the most robust is expected to be lensing extraction, because it reconstructs the matter power spectrum mainly in the linear regime. Weak lensing of CMB photons by large scale structure leads to a distorsion of the CMB maps, and creates correlations in the observed map between multipoles on different scales (unlike in a pure gaussian map). The lensing extraction technique consists in using these correlations in order to reconstruct the lensing field. In the same way, one can compute the lensing power spectrum, and get indication on the matter power spectrum over a range of scales and redshifts (typically, $1<z<3$)~\cite{Ade:2013tyw}.
Ä
CMB lensing extraction was previously advocated as a way to improve the Planck error bar on the total neutrino mass by a factor two~\cite{20-Perotto:2006rj}. This cannot happen if the two data sets (anisotropy spectrum and extracted lensing spectrum) pull the results in opposite directions. Unfortunately, this is exactly what happens with the first Planck data release. We have seen that the Planck temperature spectrum puts strong limits on the mass. Instead, the extracted lensing spectrum has a marginal preference for a non-zero $M_\nu$, because the step-like suppression induced by neutrino masses in the lensing spectrum gives a better fit. As a result, the bound gets looser when including lensing extraction: $M_\nu < 0.84$ eV (95\%; Planck+WP+highL+lensing) or $M_\nu < 0.25$ eV (95\%; Planck+WP+BAO+lensing). It will be interesting to see how these bounds will evolve in the next couple of years, with better data  and better control over systematics in the lensing extraction process. The preference for a non-zero neutrino mass in the lensing data could then go away. Instead, there is still a possibility that lensing data correctly prefers $M_\nu>0$, while some systematics in Planck data (for instance, in the low-$l$ likelihood) artificially disfavour $M_\nu \neq 0$. 

Spectral distorsions of CMB maps caused by the Sunyaev-Zel'dovitch effect allow to construct a map of galaxy clusters. The distortions provide information on the position and redshift of the clusters, and even allow to estimate their mass up to some bias factor. The data can be used to make a histogram of the number of clusters as a function of redshift, or the mass of clusters as a function of redshift. These histograms can be compared to the predictions of structure formation theory (in the mildly non-linear regime). They give another handle on the underlying matter power spectrum, and hence on neutrino masses.

The situation with Planck SZ clusters is similar to the situation with lensing: the cluster data prefers a non-zero neutrino mass, unless the bias is assumed to take extreme value, that would normally be excluded by theoretical arguments. Ref.~\cite{Ade:2013lmv} gives a combined CMB + SZ cluster result preferring a non-zero neutrino mass at the 2.5$\sigma$ level, $M_\nu = 0.22\pm0.18$ eV (95\%; Planck+WP+BAO+SZ). Ref.~\cite{Battye:2013xqa} shows how the bounds depend on several assumptions made in the SZ cluster analysis. A similar trend exists with essentially all other data on the cluster mass function. The conclusion is that something remains to be understood: either systematics are underestimated in measurements of the cluster mass function, or the Planck temperature data has some incorrect feature pushing down the neutrino masses (so far, the former assumption sounds more plausible).

The CMB data can be combined with many other probes of large scale structure (galaxy spectrum, flux spectrum of Lyman-$\alpha$ forests in quasars, weak lensing surveys). Until now, these other data sets tend to be less constraining than those discussed here \footnote{Except for a recent measurement of the amplitude of the matter power spectrum, based on the BOSS galaxy redshift survey, which leads to a preference for a non-zero neutrino mass roughly at the 3-$\sigma$ level~\cite{Beutler:2014yhv}.}.

\section{Other neutrino properties}

\subsection{Neutrino density, non-thermal distorsions and leptonic asymmetry \label{neff}}

Previous estimates of the total neutrino mass rely on the standard neutrino decoupling model, leading to a neutrino density corresponding to $N_\mathrm{eff}=3.046$ as long as neutrinos are relativistic. This model could be incorrect, or missing some physical ingredients. Hence neutrinos could in principle have a different density, coming from a different neutrino-to-photon temperature ratio, or from non-thermal distorsions. For instance, this would be the case if some exotic particles would decay after neutrino decoupling, either producing extra relic neutrinos, or reheating the thermal bath. 

Another possible reason leading to a larger $N_\mathrm{eff}$ would be the production of a very large neutrino-antineutrino asymmetry in the early universe. In that case, the phase-space distribution of neutrinos at CMB times depends on the initial leptonic asymmetry of each family, and on the efficiency of neutrino oscillations in the early universe; in any case, a leptonic asymmetry produced at early times would  tend to enhance the total neutrino density at late times.

If neutrinos were massless, all the effects described above (over- or under-production of neutrinos, non-thermal distorsions, leptonic asymmetry) would entirely be described by the value of $N_\mathrm{eff}$, because the equations of evolution of cosmological perturbations in the neutrino sector could be integrated over momentum. When neutrino masses are taken into account, this is not true anymore. In that case, $N_\mathrm{eff}$ explicitly refers to the radiation density at early times, before the non-relativistic transition of neutrinos. Still, as long as individual neutrino masses are not too large, it is useful to analyse cosmological models with free $N_\mathrm{eff}$ and $M_\nu$ set in first approximation to its minimal value (0.06 eV), or with  $N_\mathrm{eff}$ and $M_\nu$ both promoted as free parameters, with an arbitrary mass splitting. To be very rigorous, each model would require an individual analysis, but in fairly good approximation, their properties are well accounted by the two parameters $N_\mathrm{eff}$ and $M_\nu$. This is not true anymore when individual neutrino masses can be large (of the order of one or several electron-volts), as assumed in section \ref{20-sterile} on massive sterile neutrinos.

Hence it is crucial to measure $N_\mathrm{eff}$ in order to check whether we correctly understand neutrino cosmology. However, if a value of  $N_\mathrm{eff}$ larger than three was measured in CMB or LSS data, we still would not know if this come from physics in the neutrino sector, or from other relativistic relics; even assuming that such relics do not exist, we would not be able to discriminate between the different cases (shift in temperature, asymmetry or non-thermal distorsions). However, we could learn something more from other cosmological probes, such as the study of BBN, leptogenesis and baryogenesis, etc., or from laboratory experiments. For instance, we know from a joint analysis of BBN and neutrino oscillation data that in order to be compatible with measurements of primordial element abundances, the leptonic asymmetry cannot enhance the neutrino density (at CMB and current time) above $N_\mathrm{eff}\simeq3.1$ \cite{20-Mangano:2011ip}.

The CMB is sensitive to $N_\mathrm{eff}$, first, through the time of equality: when this parameter increases while all other parameters are kept fixed, equality is postponed. However, this effect can be cancelled by increasing simultaneously the matter density, by exactly the same amount. Actually, one can also renormalise the cosmological constant in order to preserve all characteristic redshifts in the evolution of the universe (radiation/matter equality, matter/$\Lambda$ equality). This transformation leaves the CMB almost invariant, but not quite. Indeed, a global increase of all densities goes with an increase of $H_0$, that is visible on small angular scale: it enhances the Silk damping effect (i.e., the effect of diffusion just before photon decoupling, important for $l>1000$). Generally speaking, a good way to describe the main effect of $N_\mathrm{eff}$ on the CMB is to say that it changes the scale of Silk damping relative to the scale of the sound horizon at decoupling. Concretely, this will appear as a shift in the damping envelope of high peaks relative to the position of the first peak \cite{20-NuCosmo,Hou:2011ec}. This effect is not the only one remaining when all densities are equally enhanced: more neutrinos also imply stronger gravitational interactions between photons and free-streaming species before decoupling. This is especially important for scales that just crossed the Hubble scale during radiation domination. The result is a small shift and damping of the acoustic peak (``baryon drag'' effect \cite{20-NuCosmo,Bashinsky:2003tk}). 

As a result of all these effects, for a minimal 7-parameter model ($\Lambda$CDM + $N_\mathrm{eff}$), the CMB data alone gives $N_\mathrm{eff}=3.36_{-0.64}^{+0.68}$ (95\%; Planck+WP+highL), well compatible with the standard prediction $N_\mathrm{eff}=3.046$. An analysis with a free lensing amplitude $A_L$ of the CMB power spectrum is described in \cite{Said:2013hta}.

Like for neutrino masses, data on the cosmological expansion at late time allows to tighten the bounds on $N_\mathrm{eff}$, because it removes the partial degeneracy observed in the CMB when all densities are increased in the same proportions. Data on BAO, supernovae or direct measurements of $H_0$ tighten the constraints on $H_0$ and/or on the matter density, and forbid to increase $N_\mathrm{eff}$ with fixed redshifts for the two equalities (radiation/matter equality, matter/$\Lambda$ equality). It is worth noticing that unlike models with a free neutrino mass, models with a free $N_\mathrm{eff}$ have enough freedom for relaxing the tensions between different data sets. For instance, in the $\Lambda$CDM + $N_\mathrm{eff}$ model, the bound on the expansion rate is $H_0=69.7_{-5.3}^{+5.8}$~km~s$^{-1}$Mpc$^{-1}$ (95\%; Planck+WP), well compatible with $H_0=73.8\pm4.8$~km~s$^{-1}$Mpc$^{-1}$ (95\%; HST) \cite{Riess:2011yx}. Even when including BAO data, one gets $H_0=69.3_{-3.4}^{+3.5}$~km~s$^{-1}$Mpc$^{-1}$ (95\%; Planck+WP+highL+BAO), still compatible with HST measurements.

From the previous discussion on the effect of $N_\mathrm{eff}$ on the CMB, one can easily infer that there is a positive correlation between measured values of $N_\mathrm{eff}$ and $H_0$. Hence, when Planck data is combined with HST data, which favours high values of the expansion rate, one gets more than 2$\sigma$ evidence for enhanced radiation, $N_\mathrm{eff}=3.62_{-0.48}^{+0.50}$ (95\%; Planck+WP+highL+HST). Instead, with BAO, the evidence disappears, $N_\mathrm{eff}=3.30_{-0.51}^{+0.54}$ (95\%; Planck+WP+highL+BAO). Since models with a free $N_\mathrm{eff}$ relax the tension between the different data sets, it makes sense to combine CMB data with BAO and HST at the same time, which gives $N_\mathrm{eff}=3.52_{-0.45}^{+0.48}$ (95\%; Planck+WP+highL+BAO+HST), 
slightly more than 2$\sigma$ evidence for enhanced radiation (see also \cite{Zheng:2014dka}).

The conclusion is that if HST data are robust, then one should consider seriously the possibility that $N_\mathrm{eff}$ exceeds the standard value, because this is one of the simplest way to relax the constraint between HST and other data sets\footnote{Another way is to assume phantom dark energy with $w<-1$ \cite{Cheng:2013csa}, which is much more difficult to motivate on a theoretical basis.}. This excess could be caused by several effects (leptonic asymmetry, non-standard neutrino phase space distribution, or any type of relativistic relics). If instead we assume that HST results are biased by systematics and should not be included, then the evidence for $N_\mathrm{eff}>3.046$ becomes very weak (although a high tensor-to-scalar ratio, like the one suggested by the very recent BICEP2 results \cite{Ade:2014xna}, also tends to slightly favour an excess in $N_\mathrm{eff}$ \cite{Giusarma:2014zza}).

\subsection{Non-standard interactions}

Non-standard neutrino interactions (beyond the weak force) could occur in many extensions of the standard model, and could affect cosmological observables in several different ways. For instance, if non-standard neutrino-electron interactions exist, it was shown that they could enhance $N_\mathrm{eff}$ at most up to $3.1$ \cite{Mangano:2006ar}.
Many other cases have been studied in the literature (see the list of references in section 5.3.4 of \cite{20-NuCosmo}), still not covering all possibilities. A few representative cases have been studied recently in \cite{Archidiacono:2013dua}, with two different assumptions concerning the type of self-interaction experienced in the neutrino sector. When the self-interaction is very efficient, neutrinos tend to behave like a relativistic fluid, instead of free-streaming particles with anisotropic stress. Strongly interacting neutrinos with vanishing anisotropic stress are ruled out by Planck data with huge significance. The best fit occurs for standard decoupled neutrinos, and the limits found by \cite{Archidiacono:2013dua} on the self-interaction cross-section are very stringent.

\subsection{Sterile neutrinos\label{20-sterile}}

There are lots of good reasons to postulate the existence of sterile neutrinos: to explain the origin of neutrino masses, the existence of neutrino oscillations, or the mechanism responsible for baryogenesis. Sterile neutrinos could also play the role of dark matter. However, the kind of sterile neutrinos that can be probed with CMB and LSS observations are light sterile neutrinos, with a mass in the eV range. These sterile neutrinos are not motivated by fundamental physics issues, but rather by a few anomalies in short baseline neutrino oscillation data (LSND, MiniBooNE, reactor experiments).

Sterile neutrinos can be generated through various mechanisms, including resonant or non-resonant oscillations with the flavour neutrinos produced in the thermal bath. Depending on the active-sterile neutrino mixing angle(s), sterile neutrinos could thermalise or not \cite{Mirizzi:2013kva}. When sterile neutrinos are in the relativistic regime, they can be entirely described in terms of an enhanced $N_\mathrm{eff}$ (close to four with one generation of thermalised sterile neutrinos). When they become non-relativistic, the parameterisation of the model is not trivial anymore: in principle, one should specify the phase-space distribution of the sterile neutrino, and vary as many mass parameters as there are neutrino species (3+1 in the simplest scenarios). Since sterile neutrinos have a significant mass, different models with the same ($N_\mathrm{eff}$, $M_\nu$) parameters but with different mass splittings and/or a different phase space distribution for the relic sterile neutrino can have distinct signatures on the CMB. 

The Planck collaboration has released results for the simple case of two active neutrinos with negligible mass, plus one active neutrino with $m=0.06$ eV, plus finally one sterile neutrino with a mass $m_s$ and a phase-space distribution equal to the one of active neutrinos multiplied by a suppression factor $\chi_s$. This corresponds to the so-called Dodelson-Widrow (DW) scenario \cite{Dodelson:1993je}, in which sterile neutrinos are generated through non-resonant oscillations. This case can be remaped into the one of thermally distributed sterile neutrinos with a smaller temperature than active neutrinos. The analysis could be carried out in terms of two free parameters ($N_\mathrm{eff}$, $m_s$), but it is more convenient to use ($N_\mathrm{eff}$, $\omega_s$), where $\omega_s \equiv \Omega_s h^2$ is the relic density of sterile neutrinos. Indeed, $\omega_s$ is the parameter really probed by the CMB through the lensing and shift-in-the-peak effects. The Planck paper reports constraints on the quantity $94.1\,\omega_s$ eV, called the effective sterile neutrino mass, because it would coincide with the real mass if the sterile neutrino distribution was the same as for active neutrinos in the instantaneous decoupling limit. Ref.~\cite{Ade:2013zuv} provides  joint constraints on ($N_\mathrm{eff}$, $94.1\,\omega_s$ eV) from CMB+BAO data, showing that a thermalised neutrino with the same temperature as active neutrinos could have a mass of at most half an electron-volt, while a DW neutrino with 1 eV mass should have  $\chi_s<0.5$. The results show no evidence at all for sterile neutrinos. 

In the previous sections, we have seen that CMB anisotropy data alone prefers a vanishing neutrino mass, and is well compatible with the standard prediction $N_\mathrm{eff}=3.046$. Instead, HST data pushes for extra radiation, while lensing extraction and SZ clusters push for a non-zero neutrino mass. When all these data are considered at the same time, one gets marginal evidence for both $N_\mathrm{eff}>3$ and $M_\nu>0$, which could be interpreted in terms of light sterile neutrinos. Refs.~\cite{Giusarma:2014zza,Wyman:2013lza,Hamann:2013iba} performed a joint analysis of all these data sets for a model with three active neutrinos of negligible mass, and one massive DW sterile neutrino. They find marginal evidence for such a sterile neutrino, with a mass in the range $m_s=0.59\pm 0.38$~eV (95\% C.L.) and with $\chi_s=0.61\pm0.60$ (95\% C.L.). These results are in moderate tension with the 3+1 scenario that could explain the short baseline anomaly (see also 
\cite{Hamann:2011ge,Gariazzo:2013gua}). Note that these results do not only apply to sterile neutrinos and can be easily transposed to the case of other light massive relics, such as thermalised axions (see e.g.\ \cite{Giusarma:2014zza,Raffelt:2010zz,Archidiacono:2013cha}).
 
\section{Sensitivity of future experiments to neutrino parameters}
\label{20-sec:future}

The next releases of the Planck CMB satellite will lead to better bounds on neutrino masses. The forecast presented in \cite{20-Perotto:2006rj} predicts a neutrino mass sensitivity of $\sigma(M_\nu) \sim 0.1$ eV from Planck alone, using full temperature and polarisation data, and lensing extraction. However, if the small internal tension between temperature and lensing found in the first release survives in the next ones, the final error bar will remain larger.

Several galaxy surveys with better sensitivity and larger volume are about to release data or have been planned over the next decades, including the Baryon Oscillation Spectroscopic Survey (BOSS), the Dark Energy Survey (DES), the Large Synoptic Survey Telescope (LSST) or the Euclid satellite. Concerning cosmic shear surveys, spectacular improvements are expected from Pan-STARRS, or the DES, LSST and the Euclid surveys already mentioned above.

In a near future, the prediction of ref.\ \cite{20-Sekiguchi:2009zs} is that the combination of full Planck data with BAO scale information from the full BOSS survey could lower the error down to $\sigma(M_\nu) \sim 0.06$ eV. In addition, the authors of \cite{20-Gratton:2007tb} find that adding Lyman alpha data from BOSS should lead to comparable sensitivities, and even better results might be expected from the addition of galaxy power spectrum data from the same survey.

In ref.\ \cite{20-Lahav:2009zr} it was found that the measurement of the galaxy harmonic power spectrum in seven redshift bins by DES should lead to a sensitivity of $\sigma(M_\nu) \sim 0.06$~eV when combined with Planck data (without lensing extraction). Similar bounds were derived in \cite{20-Namikawa:2010re} for another combination of comparable experiments. This shows that at the horizon of 2015, a total neutrino mass close to $M_\nu\simeq0.1$ eV could be marginally detected at the 2-$\sigma$ level by cosmological observations. Because this value coincides with the lowest possible total mass in the inverted hierarchy scenario, the latter could start to be marginally ruled out in case the data still prefers $M_\nu=0$. 

Many papers studied the sensitivity of Euclid to the total neutrino mass. The conservative analysis of \cite{Audren:2012vy}, taking into account a theoretical error related to the difficulty to model non-linear effects on small scales, suggests an error of the order of $\sigma(M_\nu) \sim 0.03$ eV in combination with Planck data. Constraints based on the ground-based LSST should be slightly weaker \cite{20-Hannestad:2006as}. Hence, in the early 2020's, we expect that a combination of cosmological data sets could detect the total neutrino mass of the normal hierarchy scenario, $M_\nu\simeq0.05$~eV, at the 2-$\sigma$ level. If the total mass is instead close to $M_\nu\simeq0.1$~eV, it will be detected at the 4-$\sigma$ level. However, in that case, available experiments would not have enough sensitivity to distinguish between an inverted and normal hierarchy scenario with the same $M_\nu$.

Even more progress could be provided by the promising technique of 21-cm surveys.  Instead of mapping  the distribution of hydrogen atoms trough the absorption rate of photons traveling from quasars, it should be possible to observe directly the photons emitted by these atoms at  a wavelength $\lambda\simeq21$ cm from the transition
from one hyperfine level to the other. While travelling towards the observer, these photons are redshifted, and seen with a wavelength indicating the position of the emitting atoms in redshift space.
Recent theoretical progresses in this field show that using this technique, future dedicated experiments should be able to map hydrogen and hence baryonic fluctuations at very high redshift (typically $6<z<12$), and to probe the matter power spectrum deep inside the matter-dominated regime on linear scales 
\cite{20-Pritchard:2011xb,Shimabukuro:2014ava}. This field is still in its infancy, and the forecasts presented so far have to be taken with care, due to the difficulty to make a realistic estimate of systematic errors in future data sets. A
sensitivity of $\sigma(M_\nu) \sim 0.075$~eV for the combination of Planck with the Square Kilometer Array (SKA) project, or $\sigma(M_\nu) \sim 0.0075$~eV with the Fast Fourier Transform Telescope (FFTT), was found in \cite{20-Pritchard:2008wy}. However, the authors show that such impressive experiments would still fail in discriminating between the NH and IH scenario.

An eventual post-Planck CMB satellite or post-Euclid survey would also have a great potential. The forecast analysis in \cite{20-Lesgourgues:2005yv} shows that for 
a CMB satellite of next generation one could get $\sigma(M_\nu) \sim 0.03$ eV alone, thanks to a very precise reconstruction of the CMB lensing potential, while \cite{20-Wang:2005vr} discusses the potential of cluster surveys. Finally, the authors of \cite{20-Jimenez:2010ev} show how far the characteristics of an hypothetical galaxy or cosmic shear survey should be pushed in order to discriminate between two allowed NH and IH scenarios with the same total mass.

In the future, the sensitivity of CMB and LSS experiments to $N_\mathrm{eff}$ will also increase significantly. Ref.~\cite{20-Perotto:2006rj} find that using lensing extraction, the full Planck can lower the error bar down to $\sigma(N_\mathrm{eff})\sim0.3$. Refs.~\cite{20-Kitching:2008dp,20-Carbone:2010ik} find that the combination of Planck data with the Euclid galaxy survey or cosmic shear survey would give $\sigma(N_\mathrm{eff})\sim0.1$. However, Ref.~\cite{Basse:2013zua} claims that the combination of Euclid's galaxy survey, cosmic shear survey and cluster mass function measurement would give $\sigma(N_\mathrm{eff})\sim0.02$.

\section{Conclusions}

Neutrinos, despite the weakness of their interactions and their small
masses, can play an important role in cosmology that we have reviewed
in this contribution. In addition, cosmological data can be used to
constrain neutrino properties, providing information on these elusive
particles that complements the efforts of laboratory experiments. In
particular, the data on cosmological observables have been used to
bound the radiation content of the Universe via the effective number of neutrinos, including a potential extra
contribution from other relativistic particles.

But probably the most important contribution of cosmology to our
knowledge of neutrino properties is the information that provides on
the absolute scale of neutrino masses. We have seen that the analysis
of cosmological data can lead to either a bound or a measurement of
the sum of neutrino masses, an important result complementary to
terrestrial experiments such as tritium beta decay and neutrinoless
double beta decay experiments. In the next future, thanks to the data
from new cosmological experiments we could even hope to test the
minimal values of neutrino masses guaranteed by the present evidences
for flavour neutrino oscillations.  For this and many other reasons, we
expect that neutrino cosmology will remain an active research field in
the next years.

\ack
SP was supported by the Spanish grants FPA2011-22975 and Multidark Consolider CSD2009-00064
(MINECO),  and by PROMETEO/2009/091 (Generalitat Valenciana).

\end{document}